\documentclass[aps,prl,groupedaddress,twocolumn,superscriptaddress]{revtex4-1}

\usepackage{subfigure}
\usepackage{graphicx}
\usepackage{textgreek}
\usepackage{multirow}
\usepackage{braket}
\usepackage{amssymb}
\usepackage[colorlinks=true,allcolors=blue,linktocpage,hyperfootnotes=false]{hyperref}
\usepackage{amsmath}
\usepackage{mathtools}
\usepackage[dvipsnames]{xcolor}
\usepackage{soul}
\usepackage{nicefrac, xfrac}
\usepackage{svg}

\begin{document}

\newcommand{\Ne}[1]{$^{#1}$Ne$^{9+}$}
\newcommand{\C}[1]{$^{#1}$C$^{6+}$}
\newcommand{\alpaka}{\mbox{\textsc{Alphatrap}}\ }
\newcommand{\penta}{\mbox{\textsc{Pentatrap}}\ }
\newcommand{\pentano}{\mbox{\textsc{Pentatrap}}}

\title[High-precision determination of $g$ factors and masses of 20Ne9+ and 22Ne9+]{High-precision determination of $g$ factors and masses of \Ne{20} and \Ne{22}}
\author{F. Hei{\ss}e} 
\email{fabian.heisse@mpi-hd.mpg.de}
\author{M. Door}
\email{menno.door@mpi-hd.mpg.de}
\author{T. Sailer}
\author{P. Filianin}
\author{J. Herkenhoff}
\author{C. M. K{\"o}nig}
\author{K. Kromer}
\author{D. Lange}
\author{J. Morgner}
\author{A. Rischka}
\author{Ch. Schweiger}
\author{B. Tu}
\affiliation{Max-Planck-Institut f{\"u}r Kernphysik, Heidelberg, Germany}
\author{Y. N. Novikov}
\affiliation{Kurchatov Institute-PNPI, 188300 Gatchina, Russia}
\affiliation{Saint Petersburg State University, 199034 Saint Petersburg, Russia}
\author{S. Eliseev}
\author{S. Sturm}
\author{K. Blaum}
\affiliation{Max-Planck-Institut f{\"u}r Kernphysik, Heidelberg, Germany}
\date{\today}

\begin{abstract}
We present the measurements of individual bound electron $g$ factors of \Ne{20} and \Ne{22} on the relative level of $0.1\,\text{parts}$ per billion. The comparison with theory represents the most stringent test of bound-state QED in strong electric fields. A dedicated mass measurement results in $m\left(^{20}\text{Ne}\right)=19.992\,440\,168\,77\,(9)\,\text{u}$, which improves the current literature value by a factor of nineteen, disagrees by $4$ standard deviations and represents the most precisely measured mass value in atomic mass units. Together, these measurements yield an electron mass on the relative level of $0.1\,\text{ppb}$ with $m_{\text{e}}=5.485\,799\,090\,99\,(59) \times 10^{-4}\,\text{u}$ as well as a factor of eight improved $m\left(^{22}\text{Ne}\right)=21.991\,385\,098\,2\,(26)\,\text{u}$. 
\end{abstract}

\maketitle

The theory of quantum electrodynamics (QED) successfully describes a broad scope of phenomena ranging from elementary particle physics to atoms and molecules~\cite{10.1143/PTP.1.27,PhysRev.73.416,PhysRev.76.769}. Since QED is a blueprint for all quantum field theories, it is of utmost importance to verify its sophisticated calculation methods as precisely as possible~\cite{PhysRevLett.130.071801,atoms7010028}. In particular, the deviation between the experimental results and the theoretical prediction for muon $g-2$~\cite{PhysRevLett.126.141801}, muonic deuterium hyperfine splitting~\cite{PhysRevA.98.062513} as well as the fine structure anomaly in heavy muonic systems~\cite{PhysRevLett.128.203001} currently challenge our understanding of QED. That calls for additional high-precision QED tests. Furthermore, based on such comparisons fundamental constants can be extracted, e.g. the fine structure constant $\alpha$ \cite{PhysRevLett.130.071801}, the Rydberg constant $R_\infty$~\cite{PhysRevLett.107.203001,Antognini2013} as well as the electron, proton, deuteron, muon and charged pion masses~\cite{Nature2014,Alighanbari2020,Patra2020,Kortunov2021,Alighanbari23,PhysRevLett.82.711,TRASSINELLI2016583}. 

The determination of magnetic moments and Lamb shifts of bound leptons allows tests of bound-state (BS) QED under extreme conditions, such as strong electric and magnetic fields~\cite{sturm_g_2011,PhysRevLett.94.223001,PhysRevLett.95.233003,Wilfried2017, BELTRAMI1986679}. In this letter, we present a BS-QED test with highest precision by comparing state-of-the-art bound-electron $g$-factor calculations of \Ne{20} and \Ne{22} with corresponding experimental results. The electron $g$ factor is a dimensionless constant relating the electron's magnetic moment to its spin. The energy splitting of the Zeeman levels in a magnetic field $B$ is given by: $h \nu_{\text{L}} = h \frac{g}{2} \frac{e}{2 \pi m_{\text{e}}} B$~\cite{beier_g_j_2000}. Here, $\nu_{\text{L}}$ is the Larmor spin precession frequency of an electron, $h$ is the Planck's constant and $\frac{e}{m_{\text{e}}}$ is the electron's charge-to-mass ratio. 
Additionally, the free-space cyclotron frequency of an ion in the magnetic field can be expressed by $\nu_{\text{c}}=\frac{1}{2 \pi} \frac{q_{\text{ion}}}{m_{\text{ion}}} B$, where $\frac{q_{\text{ion}}}{m_{\text{ion}}}$ is the charge-to-mass ratio of the ion. Combining both equations yields
\begin{equation}
    g_{\text{exp}} = 2 \frac{\nu_{\text{L}}}{\nu_{\text{c}}}\frac{m_{\text{e}}}{m_{\text{ion}}}\frac{q_{\text{ion}}}{e} =  2 \, \Gamma \frac{m_{\text{e}}}{m_{\text{ion}}}\frac{q_{\text{ion}}}{e} \quad . \label{eq:g-factor}
\end{equation}
In this work, the $\Gamma$ measurement is performed at the \alpaka setup~\cite{sturm_alphatrap_2019}, while the mass of $^{20}\text{Ne}^{10+}$ is measured at the \penta experiment~\cite{Repp2012}, due to the insufficient precision of the current literature value. The charge ratio $\nicefrac{q_{\text{ion}}}{e}$ is a known integer number and the electron mass as well as the mass of $^{22}\text{Ne}^{9+}$ are determined by other experiments~\cite{Nature2014,huang_ame_2021}.

The mass of $^{20}\text{Ne}^{10+}$ is determined by measuring the cyclotron-frequency ratio $\mathcal{R}^{\text{CF}}=\nicefrac{\nu_{\text{c}}\left(^{20}\text{Ne}^{10+} \right)}{\nu_{\text{c}} \left(\text{\C{12}}\right) }$. Given both $\nu_{\text{c}}$ measurements are carried out in the same $B$ field, the $^{20}\text{Ne}^{10+}$ mass can be expressed by: \mbox{$m \left(^{20}\text{Ne}^{10+} \right) = \frac{10}{6}  \, m \left( \text{\C{12}} \right) / \mathcal{R}^{\text{CF}}$}. The ion masses are linked to $ m \left(^{20}\text{Ne}^{9+} \right)$ and the neutral masses via the mass of the missing electrons~\cite{tiesinga2021} and binding energies~\cite{kramida_nist_2021}, which are known to sufficient precision. 

The \penta and \alpaka experiments are both cryogenic Penning-trap setups placed in two superconducting magnets located at the Max Planck Institute for Nuclear Physics in Heidelberg. Their trap and detection electronics are cooled to the temperature of liquid helium $\left(4\,\text{K}\right)$. The highly charged ions are produced externally in Heidelberg compact electron beam ion traps (HC-EBIT)~\cite{micke_heidelberg_2018, schweiger_2019_tipebit}, transferred via room temperature beamlines to the respective trap setups and cleaned from possible contamination ions via magnetron cleaning~\cite{heise_high-precision_2019}.
\begin{figure*}[tb]
  \includegraphics[width=\linewidth]{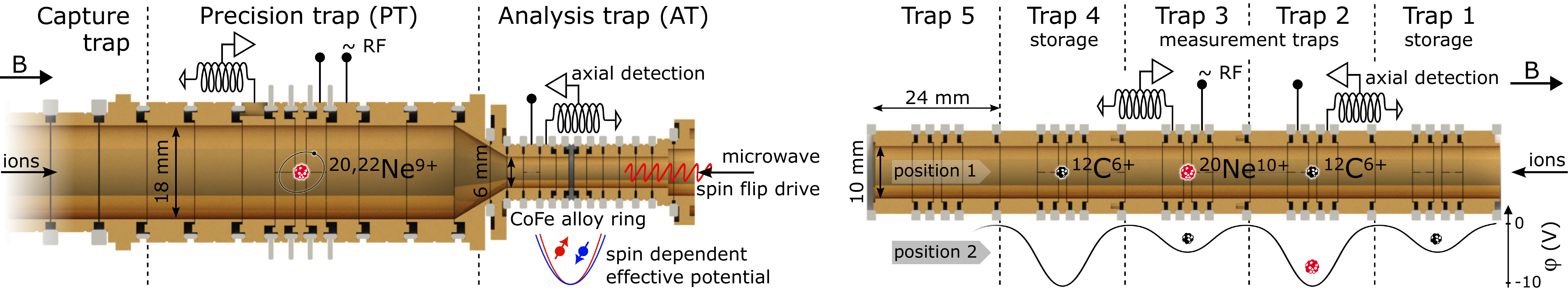}
  \caption{Sectional views of the cylindrical Penning-trap setups of \alpaka~\cite{sturm_alphatrap_2019} (left) and \penta~\cite{Repp2012} (right). Both trap towers are built from gold-plated copper ring electrodes which are isolated by sapphire and quartz rings. The connected superconducting tank circuits, the amplifiers for the axial detection systems and the RF excitation lines are shown. At the \alpaka experiment, $\nu_{\text{c}}$ is measured simultaneously to the irradiation of the microwave drive in the Precision trap. The CoFe alloy central ring electrode in the Analysis trap produces a strong magnetic bottle, allowing a non-destructive spin-state detection with close to $100\%$ fidelity. Here, a different spin orientation changes the axial force resulting from the magnetic moment inside the inhomogeneous magnetic field leading to axial frequency jumps of the ion (\Ne{20,22} yield jumps of $1.8\,\text{Hz}$ and $1.6\,\text{Hz}$, respectively)~\cite{Werth2000}. At the \penta setup, both ions' $\nu_{\text{c}}$ are measured successively in each measurement trap, resulting in one cyclotron frequency ratio per trap. The ion species are effectively swapped in the measurement traps by adiabatically transporting all ions to the respective neighbouring trap and back (position 1 and 2). The different trapping potential depths ($\mathrm{\phi}$) in the two measurement traps are due to different resonance frequencies of the axial detection circuits.}
    \label{fig:setup}
\end{figure*}
Due to the superposition of the static electric and magnetic field, the cyclotron frequency splits into three individual eigenfrequencies: the axial frequency $\nu_{\text{z}}$ on the order of hundreds of kHz, the magnetron frequency $\nu_-$ on the order of single kHz and the modified cyclotron frequency $\nu_+$ on the order of tens of MHz. Combined, these frequencies can be utilized to calculate the free-space cyclotron frequency via the invariance theorem $\nu_{\text{c}} = \sqrt{\nu_+^2+\nu_{\text{z}}^2+\nu_-^2}$~\cite{Brown1986}. Due to the strong hierarchy of the eigenfrequencies, $\nu_+$ has to be determined with highest precision, followed by $\nu_{\text{z}}$. The magnetron frequency is determined only occasionally throughout the campaign due to its stability and reduced influence on the total uncertainty on $\nu_{\text{c}}$.

In both experiments the ion's eigenfrequencies are measured non-destructively. The ion's axial oscillations induce fA image currents in high-impedance cryogenic tank circuits~\cite{DEHMELT1969109,Wineland1975} leading to a measurable voltage signal, see Fig.~\ref{fig:setup}. Interaction with the tank circuit, typically referred to as resistive cooling, causes a temperature of the axial motion near~$4\,\text{K}$. Here, the ion effectively shorts the ambient Johnson-Nyquist noise~\cite{PhysRev.32.97,PhysRev.32.110}, leading to the characteristic dip in the spectrum. The axial frequency can be directly measured by fitting the well-known dip-lineshape to the spectrum. The radial frequencies are measured via radio-frequency (RF) sideband coupling to the axial motion, resulting in a double-dip spectrum to extract the initial parameters for the more precise phase sensitive Ramsey-type measurements~\cite{PhysRevLett.63.1674,Cornell1990, Sturm2011}.

The \penta experiment ~\cite{Repp2012, roux_2012, rischka2020} is designed for high-precision mass-ratio determinations of ions, see Fig.~\ref{fig:setup}. The measurement scheme, similar to previous measurements at the \penta experiment~\cite{rischka2020,schussler2020,filianin2021,Kathrin22}, starts with three single ions of two species $\left(^{20}\text{Ne}^{10+}, ^{12}\text{C}^{6+} \right)$ loaded in alternating order into the three central traps. The traps 2 and 3 are used in parallel for the determination of $\mathcal{R}^{\text{CF}}$, while the traps 1 and 4 are exclusively used for storage. The properties of the traps are summarized in Table~\ref{tab:setup}. At \penta $\nu_{\text{z}}$ is simultaneously determined during the longest phase-evolution time of the Ramsey-type $\nu_+$ measurement~\cite{rischka2020}. This reduces the effect of electric field jitter compared to sequential measurements and allows for shorter overall measurement times. Daily reloading of the ions, as a result of the limited ion lifetime in \pentano, reduces the impact of potential eigenfrequency shifts on $\mathcal{R}^{\text{CF}}$ due to hypothetically unnoticed ion contaminations.

\begin{table}
    \caption{Trap properties of the \alpaka and \penta experiments. The magnetic field inhomogeneities are defined via $B(z) = B_0+B_1z+B_2z^2$, where $z$ is the axial position. The different $\nu_+$ for PT and AT are for \Ne{20} and $^{22}\text{Ne}^{9+}$, respectively. For \penta the corresponding eigenfrequencies of $^{20}\text{Ne}^{10+}$ and $^{12}\text{C}^{6+}$ are given.}
    \begin{ruledtabular}
    \begin{tabular}{c c c c c}
     & \multicolumn{2}{c}{\alpaka} & \multicolumn{2}{c}{\penta} \\
       & PT & AT & Trap 2 & Trap 3  \\ \hline
         $B_0 \left[ \text{T} \right]$ & $\approx 4$ & $\approx 3.9$ & \multicolumn{2}{c}{$\approx 7$}  \\
        $B_1 \left[ \frac{\text{mT}}{\text{m}} \right] $ &  $2.64\,(3)$ & $ \genfrac{}{}{0 pt}{1}{<}{>}  \pm 10^4$  &  $1.41\,(27)$ & $-1.49\,(16) $ \\
         $B_2 \left[ \frac{\text{mT}}{\text{m}^2} \right] $ & $64\,(3)$ & $43.1\,(1)\times 10^{6}$ & $64\,(5) $ & $22\,(5)$  \\ 
         $\nu_+ \left[ \text{MHz} \right] $ &  $\approx 28/25$ & $\approx 27/24$ &  \multicolumn{2}{c}{$\approx 54$}  \\
         $\nu_{\text{z}} \left[ \text{kHz} \right] $ & $\approx 651$ & $\approx 334$ & $\approx 736$ & $\approx 502$  \\
         $\nu_- \left[ \text{kHz} \right]$ & $\approx 8$ & $\approx 2$ & $\approx 5$ & $\approx 2$ \\
         $\nu_{\text{L}} \left[ \text{GHz} \right]$ & $\approx 113$ & $\approx 108$ & -- & --  \\ 
         $T_{\text{z}} \left[ \text{K} \right] $ & \multicolumn{2}{c}{$5.7\,(3)$} & \multicolumn{2}{c}{$7\,(2)$}  \\
    \end{tabular}
    \label{tab:setup}
    \end{ruledtabular}
\end{table}

\penta has an active stabilization system for the helium gas pressure and liquid helium level in the cold bore of the magnet, resulting in a very stable magnetic field. The temporal variation is dominated by a constant relative degradation of about $0.3\,\frac{\text{ppb}}{\text{h}}$. Therefore, it is sufficient to measure only one full set of unwrapping phases for different evolution times at the beginning and merely the phase of the shortest $\left(\phi^{\text{ref}}\right)$ and longest evolution time   $\left(\phi^{\text{precision}}\right)$ in all subsequent measurement cycles to determine $\nu_+$.

\begin{figure}
\centering
    \subfigure{\label{fig:pentadata_a}\includegraphics[width=0.98\linewidth]{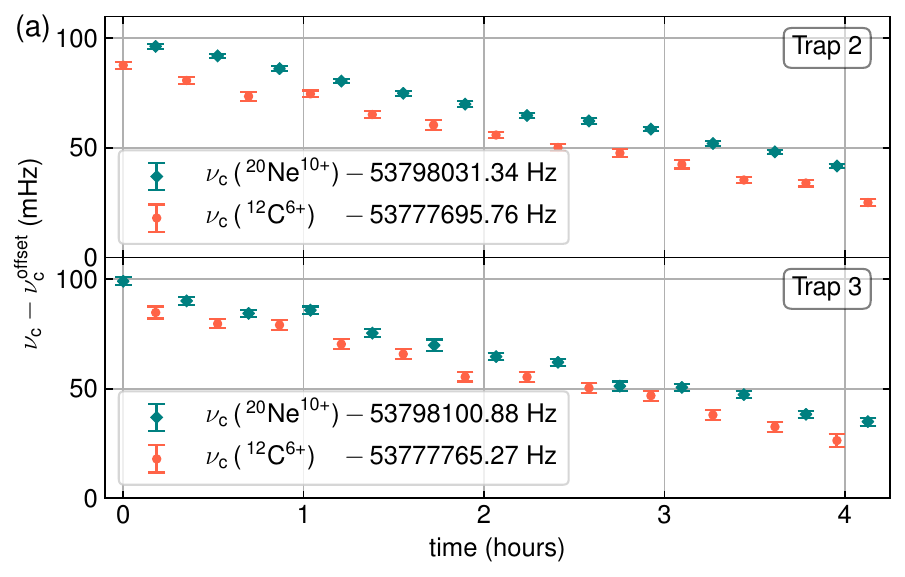}}
    \hfill
    \vspace{-16pt}
    \subfigure{\label{fig:pentadata_b}\includegraphics[width=0.97\linewidth]{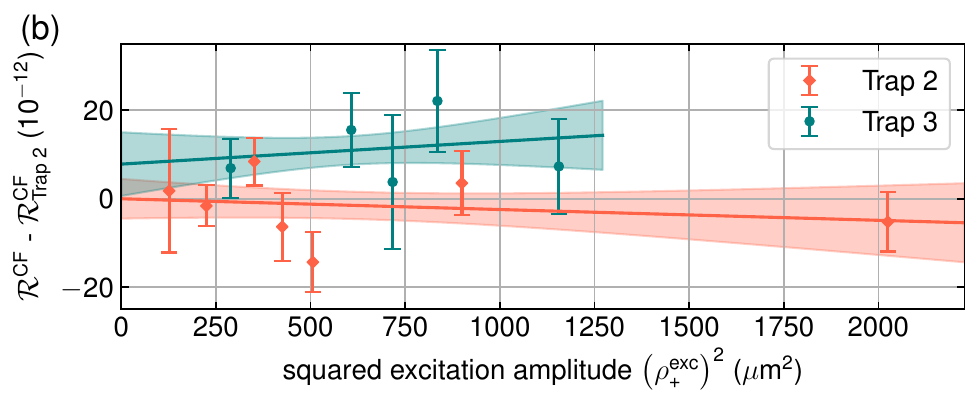}}
    \hfill
    \caption{(a) Exemplary $\nu_{\text{c}}$ data from traps 2 and 3. (b) Determined $\mathcal{R}^{\text{CF}}$ for different $\left(\rho_+^{\text{exc}}\right)^2$ including the corresponding least-square fits and $1\,\sigma$ confidence bands with $\chi_{\text{red}}^2 = 1.11$ and $0.65$ for traps 2 and 3, respectively. Higher-order shifts $\propto \left(\rho_+^{\text{exc}}\right)^3$ and $\left(\rho_+^{\text{exc}}\right)^4$ are estimated to be negligible. The $\mathcal{R}^{\text{CF}}$ of each trap is extracted from the fit at $\left(\rho_+^{\text{exc}}\right)^2 = 0 \, \mu \text{m}^2 $. This extrapolation reduces the relativistic shift and all other shifts that are proportional to $\left(\rho_+^{\text{exc}}\right)^2$.}
    \label{fig:pentadata}
\end{figure}

After the determinations of ten $\phi^{\text{ref}}$ and ten $\phi^{\text{precision}}$ the ion positions are swapped by transporting the ions to the adjacent traps and the phase measurements are repeated. Each cycle, including forty phase measurements, ion cooling and transports, takes about $22$~minutes. An example of measurement data is shown in Fig. \ref{fig:pentadata_a}. To accommodate for the slow magnetic field loss of our magnet, each cycle's $\mathcal{R}^{\text{CF}}$ is determined using interpolated data of the free cyclotron frequencies of the two ions to a common time in each trap~\cite{filianin2021}.

The statistical shot-to-shot jitter of $\mathcal{R}^{\text{CF}}$ (about $7\times 10^{-11}$) originates mainly from the ion's thermally distributed initial radius of the modified cyclotron motion, which averages io approximately $2~\,\mu\text{m}$ and takes a slightly different value after each cooling cycle. This distribution persists after the excitation pulse for the Ramsey-type measurement scheme and results in a variation of the relativistic frequency shift on $\nu_+$. These fluctuations lead to a phase jitter that increases linearly with the phase evolution time and thus limits the phase evolution time to $20\,\text{s}$. The excitation radii were varied between $\rho_+^{\text{exc}} = 15 - 50\,\mu\text{m}$ to extrapolate $\mathcal{R}^{\text{CF}}$ to virtually zero excitation for each trap~\cite{PhysRevLett.119.033001,heise_high-precision_2019}, see Fig.~\ref{fig:pentadata_b}.

All systematic effects are summarized in Table~\ref{tab:pentasys}. The main systematic uncertainties are caused by the dip-fit lineshape~\cite{rau_penning_2020}, the non-linear phase readout effect~\cite{filianin2021} and by the trapping-potential depth difference to match the two ions axial frequencies with the detection systems resonance~\cite{rischka2020}.

\begin{table}
\caption{\label{tab:pentasys} Relative systematic corrections and their uncertainties on  ${\mathcal{R}}^{\text{CF}}$ at \penta after the extrapolation to zero-excitation amplitude for both traps. All values are given in parts per trillion ($10^{-12}$). The systematic shifts are assumed to be uncorrelated between the traps.}
\begin{ruledtabular}
\begin{tabular}{lrr}
Effect (parameters) & Trap 2 & Trap 3 \\ \hline
$T_z$ ($B_2$, $C_3$, $C_4$, $C_6$) & $-0.8\,(0.3)$ & $-1.7\,(0.5)$\\
Magnetron frequency & $0.0\,(0.5) $ & $0.0\,(0.5) $\\
Non-linear phase read-out~\cite{filianin2021} & $0.0\,(1.0) $ & $0.0\,(1.9) $\\
Position difference~\cite{rischka2020} & $0.0\,(1.0)$ & $0.0\,(0.8)$ \\
Dip lineshape~\cite{rau_penning_2020} & $0.0\,(1.0)$ & $0.0\,(1.8)$ \\
Image charge shift~\cite{vandyckICS1989, vandyckICS2006, schuhICS2019} & $19.5\,(1.0)$ & $19.5\,(1.0)$ \\ \hline
Total &  $18.7\,(2.1)$ & $17.8\,(3.0)$ \\ 
\end{tabular}
\end{ruledtabular}
\end{table}

Given these effects, the determined $\mathcal{R}^{\text{CF}}$ for both traps are
\begin{equation*}
\begin{split}
    \mathcal{R}^{\text{CF}}_{\text{Trap 2}} =& 1.000\,378\,141\,801\,(5)(2) \, , \\ \label{eq:cfr20ne} 
    \mathcal{R}^{\text{CF}}_{\text{Trap 3}} =& 1.000\,378\,141\,807\,(7)(3) \, , 
    \end{split}
\end{equation*}
with the first and the second bracket showing the statistical and systematic uncertainty, respectively. The weighted mean of the two traps yields $\mathcal{R}^{\text{CF}}_{\text{mean}} = 1.000\,378\,141\,802\,(4)$.
To determine $m\left(^{20}\text{Ne}\right)$, we added the masses of the missing electrons and their corresponding binding energies~\cite{Nature2014,kramida_nist_2021} with an additional uncertainty of $2\,\times 10^{-11}\,\text{u}$ resulting in 
\begin{equation*}
m\left(^{20}\text{Ne}\right)\,=\,19.992\,440\,168\,77\,(9)\,\text{u} \, .
\end{equation*}
This represents the most precise mass value ever measured in atomic mass units and deviates by $4\,\sigma$ from the current literature value~\cite{huang_ame_2021, PhysRevLett.73.1481}. The possibility to achieve such a precision for atomic masses paves the way for order of magnitude improved $m\left(^{133}\text{Cs}\right)$ and $m\left(^{87}\text{Rb}\right)$, which is of utmost importance for the determination of $\alpha$ via photon-recoil experiments~\cite{Parker2018,Morel2020}. For the latter, the mass uncertainty currently represents the second largest contribution within the whole uncertainty budget.

The $g$-factor measurements were carried out in the \alpaka setup~\cite{sturm_alphatrap_2019,arapoglou_g_2019,egl_application_2019}. A cryo-valve separates the beamline vacuum and the trap vacuum. This ensures a vacuum of better than $10^{-16}\,\text{mbar}$ in the trap section, enabling ion lifetimes on the order of months.

The setup includes an Analysis trap (AT) enabling the detection of the electron's spin-state orientation with respect to the magnetic field via the continuous Stern-Gerlach effect~\cite{Dehmelt1989,Werth2000} as well as  a spectroscopic Precision trap (PT), distinguished by a harmonic electric and a homogeneous magnetic field, see Fig.~\ref{fig:setup}. In the PT the spin-flip transition is probed to highest precision by irradiating a microwave (MW) pulse $\nu_{\text{MW}}$, while simultaneously monitoring the magnetic field via $\nu_{\text{c}}$.

The double-trap measurement sequence is similar to~\cite{Nature2014,Koehler_2015}. Each measurement cycle begins by placing the respective single neon ion in the AT and determining the electron spin-state via microwave irradiation at the corresponding Larmor frequency of the AT, see Tab.~\ref{tab:setup}. Next, it is transported into the PT to measure $\nu_{\text{z}}$ and $\nu_+$ via the dip and double-dip method followed by the phase-senstive measurement of $\nu_+$~\cite{Sturm2011}. During the longest phase-evolution time of $\nu_+$ ($8\,\text{s}$ in this work), a MW pulse is simultaneously irradiated around the expected $\nu_{\text{L}}$. Together with $\nu_{\text{c}}$ this yields one $\Gamma_{\text{i}}=\nicefrac{\nu_{\text{MW}}}{\nu_{\text{c}}} $. Finally, the ion is transported back into the AT for the spin-state determination, resulting in a cycle-length of 25 minutes.

\begin{figure}
	\includegraphics[width=1\linewidth]{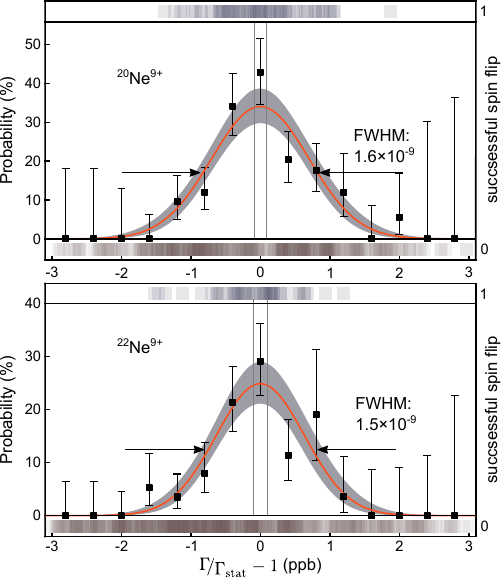}
	\caption{$\Gamma$ resonance for \Ne{20} (top) and \Ne{22} (bottom). The upper resonance contains 425 cycles, whereas the lower one contains 585 cycles. Both resonances are not power saturated and fitted by the Maximum-likelihood method with a three-parameter Gaussian lineshape (red line), since asymmetric effects shift the resonance center far below the ppt level~\cite{Brown1986,PhysRevLett.92.093002,Koehler_2015}. Here, the amplitude, the resonance width and the centre of the resonance are the free parameters. The $1-\sigma$ confidence band of the fit is displayed in grey. The black binned data points with binomial error bars are presented to guide the eye. The grey shadows at the top are observed spin flips in the PT for the respective $\Gamma$, whereas the ones at the bottom are no spin flips.}
	\label{fig:resonance}
\end{figure}

For each cycle with a spin-state change in the PT, the corresponding $\Gamma_{\text{i}}$ is assigned with $1$, whereas in the other case with $0$. This boolean information of different probed $\Gamma_{\text{i}}$'s results in a probability distribution (resonance), see Fig.~\ref{fig:resonance}. The resonance width is mainly given by the intrinsic magnetic field fluctuations during the $8~\text{s}$ probe-time of $\Gamma$.

The main systematic correction of $\Gamma\left(\text{\Ne{20}}\right)$ and  $\Gamma\left(\text{\Ne{22}}\right)$ is due to the special relativity effect caused by the excited cyclotron radii of $\rho_+^{\text{exc}} \approx 20~\mu $m for both isotopes. It amounts to a relative shift of $-5.3(11) \times 10^{-11}$. The image charge shift correction is $-2.5(1) \times 10^{-11}$~\cite{vandyckICS1989, vandyckICS2006, schuhICS2019}. Corrections and uncertainties due to magnetic inhomogeneities, electric field anharmonicities and dip lineshapes are on or below the $5\times 10^{-12}$ level.

After finishing all measurements, we identified an unexpected systematic shift originating from technical peculiarity of the cryogenic RF switches during the phase sensitive detection of $\nu_+$. Test measurements set an upper limit of a potential systematic shift of $\Gamma$ at relative $3.0\times 10^{-11}$, for more details see~\cite{Tim_2022}. This effect is diminished by not using the switches for future measurements. The final results of both $\Gamma$(\Ne{20,22}), corrected for systematic effects, are summarized in Table~\ref{tab_gfactor} and are limited by statistics. 

\begin{table}
    \centering
    \caption{The measured $\Gamma$'s with the \alpaka apparatus for \Ne{20,22} are given including their statistic and systematic uncertainties. Measured and theoretical $g$ factors are shown with combined uncertainties. For the $g$ factors the current CODATA $m_{\text{e}}$, the AME $m\left(^{22}\text{Ne}\right)$ and $m\left(^{20}\text{Ne}\right)$ determined in this work are used. The individual BS-QED  $g$-factor contributions are only listed below  $g\left(\text{\Ne{20}} \right)$ since they are equal for both isotopes. The nuclear size and recoil contribution are listed individually. The BS 2-loop uncertainties of $\mathcal{O}\left(\left(Z \alpha \right)^{6}\right)$ are dominating.}
    \label{tab_gfactor}
    \begin{ruledtabular}
    \begin{tabular}{l r r}
    & \multicolumn{1}{c}{$^{22}\text{Ne}^{9+}$}     &  \multicolumn{1}{c}{$^{20}\text{Ne}^{9+}$} \\
    \hline
    $\Gamma$  & $4450.460\,396\,87\,(42)(14) $ &  $4045.837\,341\, 56\,(34)(13)$  \\
     $g_{\text{exp}}$  &  $1.998\,767\,262\,20 (174)$ & $1.998\,767\,276\,99 (19)$   \\
      $g_{\text{theo}}$ & $1.998\,767\,263\,64 (12)$ & $1.998\,767\,277\,11 (12)$  \\
    \hline
     \hline
\multicolumn{2}{l}{Dirac (relativistic point nucl.)~\cite{Breit28}} & 1.996\,445\,170\,90 \hspace{0.46cm}   \\ 
    \multicolumn{2}{l}{Free-electron QED~\cite{tiesinga2021}} & 0.002\,319\,304\,35 \hspace{0.46cm} \\
    \hline
   \multicolumn{2}{l}{BS 1-loop Self energy~\cite{PhysRevA.69.052503,PhysRevA.95.060501,Yerokhin_2013}}  & 0.000\,002\,717\,05 \hspace{0.46cm}  \\
\multicolumn{2}{l}{BS 1-loop Vacuum polarisation~\cite{sailer_measurement_2022}}  & -\,0.000\,000\,063\,22 \hspace{0.46cm}  \\
    \hline
  \multicolumn{2}{l}{BS 2-loop~\cite{sailer_measurement_2022}} & -\,0.000\,000\,003\,17(12)   \\
     \hline
     \multicolumn{2}{l}{BS $3$-loop~\cite{PhysRevLett.24.39}} & 0.000\,000\,000\,03 \hspace{0.46cm}  \\
     \hline
     Recoil~\cite{sailer_measurement_2022}  &  0.000\,000\,133\,10 \hspace{0.46cm} & 0.000\,000\,146\,41 \hspace{0.46cm}  \\
     Size~\cite{sailer_measurement_2022}  & 0.000\,000\,004\,60 (1)  \hspace{0.001cm} &  0.000\,000\,004\,76 (1) \hspace{0.001cm} \\    
    \end{tabular}
    \end{ruledtabular}
\end{table}

The comparison of the experimental $g \left(\text{\Ne{20}}\right)$ with theory~\cite{sailer_measurement_2022} yields the most stringent test of BS-QED in strong electric fields to date at a level of $84~\text{ppm}$, slightly surpassing the test via $g \left(^{12}\text{C}^{5+}\right)$ and $g  \left(^{28}\text{Si}^{13+}\right)$~\cite{Nature2014,sturm_g-factor_2013,PhysRevLett.120.043203,PhysRevA.96.012502}.

The QED binding corrections to the free-electron $g$ factor are calculated in the $\alpha$ expansion and include one-, two- and three-loop contributions~\cite{beier_g_j_2000}. The one-loop correction is calculated in all orders of $Z \alpha$, whereas for the two-loop correction the perturbation series expansion has been calculated up to $\mathcal{O} \left( \left(Z \alpha  \right)^5\right)$~\cite{PhysRevA.102.050801}. The estimated uncertainty from the uncalculated terms of $\mathcal{O} \left(\left(Z \alpha \right)^{6}\right)$ is currently limiting all $1s~g$-factor calculation for elements with $Z\geq 4$~\cite{PhysRevA.72.022108,PhysRevLett.120.043203}. The three-loop correction is currently calculated up to $\mathcal{O}\left(\left(Z \alpha\right)^2\right)$~\cite{PhysRevLett.24.39}. The relative theory precision of $g \left(\text{\Ne{20}} \right)$ is $6 \times 10^{-11}$ and comparable with our experimental uncertainty.

In this work, the one-loop self-energy corrections are tested at an unprecedented level, verifying the corresponding calculations ~\cite{PhysRevA.95.060501,Yerokhin_2013}. In contrast, the vacuum polarisation contributions are best tested in muonic atoms~\cite{BELTRAMI1986679}. Here, the two-loop contributions are tested at the level of $7\,\%$, confirming the $Z \alpha$ expansion for intermediate-$Z$. Additionally, the total nuclear recoil contribution of the $g$ factor is precisely verified. This is complementary to the $\Delta g$ measurement~\cite{sailer_measurement_2022}, where the differential nuclear recoil contributions have been confirmed without testing any QED correction calculated within the $Z \alpha$ expansion.

Although the previously measured $g_{\text{exp}}\left(^{28}\text{Si}^{13+}\right)$ is more precise compared to $g_{\text{exp}}\left(^{20}\text{Ne}^{9+}\right)$, the uncalculated  $\mathcal{O} \left( \left(Z \alpha \right)^{6} \right) $ contributions of $g_{\text{theo}}\left(^{28}\text{Si}^{13+}\right)$ limit a more stringend QED test in that case. Tests based on the measured $g\left(^{12}\text{C}^{5+}\right)$~\cite{Nature2014} are possible via an independent $m_{\text{e}}$ determination based on the ro-vibrational transition energies of $\text{HD}^+$\cite{Alighanbari2020,Patra2020,Kortunov2021,Alighanbari23,doi:10.1080/00268976.2023.2216081} combined with the proton's and deuteron's atomic masses~\cite{huang_ame_2021,PhysRevLett.127.243001}. In fact the precision of $g_{\text{exp,theo}}\left(^{12}\text{C}^{5+}\right)$ and $m_{\text{e}}\left(\text{HD}^+\right)$ is higher compared to $g \left(^{20}\text{Ne}^{9+}\right)$, but the corresponding BS-QED contributions are reduced as well due to their $Z \alpha$ scaling.

Our presented value together with the so far measured high-precision $g$ factors ($^{12}\text{C}^{5+}$, $^{28}\text{Si}^{13+}$ and $^{118}\text{Sn}^{49+}$~\cite{Jonathan23}) constrain uncalculated higher order contributions and eventually  experimentally enhance them in the future~\cite{PhysRevLett.120.043203}.

Additionally, the agreement of the experimental and theoretical $g$ factors provides a consistency check for $m\left(^{20}\text{Ne}\right)$ determined in this work, which deviates from the literature value~\cite{huang_ame_2021}.
Furthermore, this agreement confirms the limits set by $g\left(^{28}\text{Si}^{13+}\right)$ on the $y_{\text{e}}y_{\text{n}}$ coupling constant present in the Higgs-relaxion mixing scenario~\cite{DEBIERRE2020135527,PhysRevA.106.062801}. Together with the direct $\Delta g$ measurement~\cite{sailer_measurement_2022} these $g$-factor results yield $\left[g \left(\text{\Ne{20}} \right) - g \left(\text{\Ne{22}} \right) \right]-  \Delta g = 1.3 (1.7)\times 10^{-9} $. This could be interpreted as a successful 1-2-3 test at the $200\,\text{Hz}$ level ($1.7\times10^{-9} \times 113\,\text{GHz}$) in the microwave regime to test quantum mechanics~\cite{PhysRevA.90.042102, PhysRevA.94.042117, PhysRevA.106.032209}. 

Additionally, Eq.~(\ref{eq:g-factor}) can be rearranged to express the electron mass via the theoretical value $g_{\text{theo}}$:
\begin{equation}
    m_{\text{e}} = \frac{1}{2} g_{\text{theo}} \frac{m_{\text{ion}}}{\Gamma} \frac{e}{q_{\text{ion}}} = 5.485\,799\,090\,99 \,(59) \times 10^{-4} \, \text{u} \, , \label{eq:e_mass}
\end{equation}
which is limited by the uncertainty of $\Gamma \left(^{20}\text{Ne}^{9+}\right)$. Our result is in agreement with the current CODATA value $\left(0.6\,\sigma\right)$ and a factor of three less precise~\cite{tiesinga2021}. 

If we solve~Eq.~(\ref{eq:g-factor}) for $m\left(\text{\Ne{22}}\right)$ and correct for the masses of the missing electrons~\cite{tiesinga2021} and the corresponding binding energies~\cite{kramida_nist_2021} the result yields a factor of eight improved mass uncertainty:
\begin{equation*}
m\left(^{22}\text{Ne}\right)  = 21.991\,385\,098\,2\,(26)\,\text{u} \, , 
\end{equation*}
which is in accordance with the literature value ($0.8\,\sigma$)~\cite{huang_ame_2021}.

In conclusion we determined two experimental $g$ factors of $^{20,22}$Ne$^{9+}$ and the atomic mass of $^{20}$Ne to highest precision. Together with the corresponding theoretical $g$ factors this yields the most precise test of BS-QED in strong electric fields. The atomic masses of $^{20,22}$Ne are improved by a factor of sixteen and eight, respectively. Additionally, the electron mass is determined with a relative precision of $1 \times 10^{-10}$.

This work is supported by the Max-Planck-Gesellschaft (MPG), the International Max-Planck Research Schools for Precision Tests of Fundamental Symmetries (IMPRS-PTFS) and for Quantum Dynamics in Physics, Chemistry and Biology (IMPRS-QD). The project received funding from the European Research Council (ERC) under the European Union’s Horizon 2020 research and innovation programme under Grant agreement number 832848 - FunI and by the DFG (German Research Foundation) – Project-ID 273811115 – SFB 1225 ISOQUANT. Furthermore, we acknowledge funding and support by the Max Planck, RIKEN, PTB Center for Time, Constants and Fundamental Symmetries. This work comprises parts of the PhD thesis work of M.D. to be submitted to Heidelberg University, Germany. F.H., M.D. and T.S. contributed equally to this work.

\end{document}